\documentclass[showpacs,preprintnumbers,amsmath]{revtex4}
\usepackage{graphicx}
\usepackage{epsfig}
\usepackage{bm}

\makeatletter \leftline{\epsfbox{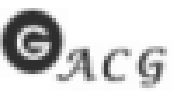}}
 \leftline
{\footnotesize{\textbf{\textsf{GACG-04/07}}}}

\newcommand{\be}{\begin{equation}}
\newcommand{\en}{\end{equation}}
\newcommand{\bea}{\begin{eqnarray}}
\newcommand{\ena}{\end{eqnarray}}

\begin{document}

\title{Open inflationary universes in a brane world cosmology}
\author{Sergio del Campo, Ram\'on Herrera and Joel Saavedra}
\address{Instituto de F\'{\i}sica, Pontificia Universidad
Cat\'olica de Valpara\'{\i}so, Casilla 4950, Valpara\'{\i}so.}
\date{\today}

\begin{abstract}
In this paper, we study a type of one-field model for open
inflationary universe models in the context of the brane world
models. In the scenario of a one-bubble universe model, we
determine and characterize the existence of the Coleman-De Lucia
instanton, together with the period of inflation after tunneling
has occurred. Our results are compared to those found in the
Einstein theory of Relativistic Models.
\end{abstract}

\maketitle

\section{\label{sec:level1} Introduction}

Recent observations from the WMAP~\cite{wmap} are entirely
consistent with a universe having a total energy density that is
very close to its critical value, where the total density
parameter has the value $\Omega =1.02\pm 0.04$. Most people
interpret this value as corresponding to a flat universe. But,
according to this result, we might take the alternative point of
view of having a marginally open or close universe
model~\cite{ellis-k}, which early presents an inflationary period
of expansion. This approach has already been considered in the
literature,~\cite{re6}\cite{re8} or \cite{shiba}\cite{ramon}, in
the context of the Einstein theory of relativity or a
Jordan-Brans-Dicke (JBD) type of gravitation theory, respectively.
All these models have been worked out in a four dimensional
spacetime. At this point, we should mention that Ratra and Peebles
were the first to elaborate on the open inflation model
\cite{Ratra}.

The idea of considering an extra dimension has received great
attention in the last few years, since it is believed that these
models shed light on the solution to fundamental problems when the
universe is traced back to a very early time. Specifically, the
possibility of creating an open or closed universe from the
context of a Brane World (BW) scenario \cite{MBPGD}, has quite
recently been considered.

BW cosmology offered a novel approach to our understanding of the
evolution of the universe. The most spectacular consequence of
this scenario is the modification of the Friedmann equation, in a
particular case when a five dimensional model is considered, and
where the matter described through a scalar field, is confined to
a four dimensional Brane, while gravity can be propagated in the
bulk. These kinds of models can be obtained from a higher
superstring theory~\cite{witten}. For a comprehensible review of
BW cosmology, see~\cite{lecturer} for example. Specifically,
consequences of a chaotic inflationary universe scenario in a BW
model was described~\cite{maartens}, where it was found that the
slow-roll approximation is enhanced by the modification of the
Friedmann equation. The purpose of the present paper is to study
an open inflation universe model, where the scalar field is
confined to the four dimensional Brane.

The plan of the paper is as follows: In Sec. II we specify the
effective four dimensional cosmological equations from a five-AdS
BW model. We write down the field equations in a Euclidean
spacetime, and we solve them numerically. Here, the existence of
the Coleman-De Lucia (CDL) instanton for two different models are
described. In Sec. III we determine the characteristic of an open
inflationary universe model that is produced after tunneling has
occurred. In Sec. IV we determine the corresponding density
perturbations for our models. In any case, our results are
compared to those analogous results obtained by using Einstein's
theory of gravity. Finally, we conclude in Sec. V.


\section{\label{sec:level2}The Euclidean cosmological equations in
Randall-Sundrum type II Scenario}

We start with the action given by
\begin{equation}
\hspace{0cm}\displaystyle S\,=M_{5}^{3}\int {d^{5}x\,\sqrt{-G}}\,\,\left(
^{(5)}\,R-2\Lambda _{5}\right) -\,\int d^{4}x{\sqrt{-g}\mathcal{L}}_{matter},
\label{ac1}
\end{equation}
where
\[
{\mathcal{L}}_{matter}(\phi )\,=\,\frac{1}{2}\partial _{\mu }\phi
\partial ^{\mu }\phi \,+\,V(\phi ),
\]
describes the matter confined in the brane $^{(5)}R$ is the Ricci
scalar curvature of the metric $G_{ab}$ of the five dimensional bulk; $M_{5}$ and $%
\Lambda _{5}$ denote the five-dimensional Planck mass and
cosmological constant, respectively. The following relations are
found to be valid in this case \be
\Lambda _{4} = \frac{4\pi }{M_{5}^{3}}(\Lambda _{5}+\frac{4\pi }{3M_{5}^{3}}%
\sigma ^{2}),  \label{relation1} \en and \be M_{4}
=\sqrt{\frac{3}{4\pi }}(\frac{M_{5}^{2}}{\sqrt{\sigma }})M_{5},
\label{mass}
\en where $\Lambda _{4}$ represents the effective cosmological
constant on to brane and $\sigma $ corresponds to the brane
tension.

\smallskip For this theory, Shiromizu et al.\cite{maeda} have shown that the
four-dimensional Einstein equations induced on the brane can be
written as \be G_{\mu \nu }=-\Lambda _{4}g_{\mu \nu }+(\frac{8\pi
}{M_{4}^{2}})T_{\mu \nu }+(\frac{8\pi }{M_{5}^{2}})S_{\mu \nu
}-\mathcal{E}_{\mu \nu }, \label{inducedeq} \en where $T^{\mu \nu
}$ is the energy-momentum tensor of the matter in the brane;
$S_{\mu \nu }$ is the local correction to standard Einstein
equations due to the extrinsic curvature; and $\mathcal{E}_{\mu
\nu }$ is the nonlocal effect corrections from a free
gravitational field, which arises from the projection of the bulk
Weyl tensor. These quantities are given by
\begin{equation}
S_{\mu \nu }=\frac{1}{12}\rho ^{2}u_{\mu }u_{\nu }+\frac{1}{12}(\rho
+2p)(g_{\mu \nu }+u_{\mu }u_{\nu }),  \label{cuadraticterm}
\end{equation}
where $\rho $ and $p$ represent the energy density and pressure of
a fluid, respectively, and
\begin{equation}
\mathcal{E}_{\mu \nu }=(\frac{M_{4}^{2}}{8\pi })^{2}\left( \frac{A}{a^{4}}%
u_{\mu }u_{\nu }+\pi _{\mu \nu }\right) ,  \label{weyl}
\end{equation}
where $A$ is a constant, and $\pi _{\mu \nu }$ is the anisotropic
stress. Since we are considering an AdS$_{5}$ bulk and a
Friedman-Robertson-Walker (FRW) brane, we should have $\pi _{\mu
\nu }=0.$ On the other hand, an extended version of Birkhoff's
theorem tells us that if the bulk spacetime is AdS, then
$\mathcal{E}_{\mu \nu }=0$ \cite{lecturer,Bowcock:2000cq}.
Finally, we can put $\Lambda _{4}=0,$ when a finned tuning is made
over $\Lambda _{5}$.

In order to write down the field equations, recall that, due to
Bianchi identity, i.e. $\nabla ^{\mu }G_{\mu \nu }=0$, we have
\begin{equation}
(\frac{8\pi }{M_{4}^{2}})\nabla ^{\mu }T_{\mu \nu }+(\frac{8\pi }{M_{5}^{2}}%
)\nabla ^{\mu }S_{\mu \nu }-\nabla ^{\mu }\mathcal{E}_{\mu \nu
}=0. \label{bianchi1}
\end{equation}
The first term is automatically satisfied and thus the following
constraint is thus obtained
\begin{equation}
\nabla ^{\mu }\mathcal{E}_{\mu \nu }=(\frac{8\pi
}{M_{5}^{2}})\nabla ^{\mu }S_{\mu \nu }\,.  \label{constraint}
\end{equation}
Now, since in our case we have $\mathcal{E}_{\mu \nu }= 0$,
eq.~(\ref{constraint}) becomes~\cite{maartens}
\begin{equation} \nabla ^{\mu
}S_{\mu \nu }=0.  \label{cosntraint2}
\end{equation}

The $O(4)$ invariant Euclidean spacetime metric is written as
\begin{equation}
\displaystyle d{s}^{2}\,=\,d{\tau }^{2}\,+\,a(\tau )^{2}\,(\,\,d{\psi }%
^{2}\,+\,\sin ^{2}\psi \,d{\Omega _{2}^{2}}\,\,).  \label{met}
\end{equation}

The scalar field equation becomes
\begin{equation}
\displaystyle \phi ''+3\frac{a'}{a}\phi'-V_{,\phi }(\phi )=0\,\,,
\label{ecphi}
\end{equation}
where $a(\tau )$ is the scale factor, and the prime represents a
derivative with
respect to the Euclidean time ($\tau $) and $V_{,\phi }(\phi )=\frac{dV}{%
d\phi }$.

When the metric~(\ref{met}) is introduced into equation
(\ref{inducedeq}), we obtain the following field equations for
scalar factor:
\begin{equation}
\displaystyle \left( \frac{a'}{a}\right)
^{2}=\frac{1}{a^{2}}-\frac{8\pi }{3M_{4}^{2}}\rho _{E}\left(
1+\frac{\rho _{E}}{2\sigma }\right) , \label{ec2}
\end{equation}
where $\rho _{E}$ corresponds to
Euclidean energy density associated with the scalar field, $\rho _{E}=-\frac{1}{2}%
\phi'^{2}+V(\phi )$. From now on we will use units where $%
M_{4}=G_{4}^{-1/2}=1$ and $\kappa =8\pi \,/M_{4}^{2}=8\pi $.

From equations (\ref{ecphi}) and (\ref{ec2}) we obtain
\begin{equation}
\displaystyle \frac{a''}{a}=-\frac{8\pi }{3M_{4}^{2}}\left(
\phi'^{2}+V(\phi )+\frac{1}{8\sigma }(5\phi'^{2}+2V(\phi
))(-\phi'^{2}+2V(\phi ))\right) . \label{ec5}
\end{equation}

We consider the effective scalar potential, $V(\phi )$,  to be of
the form analogous to that described in Ref.~\cite{ramon}:
\begin{equation}
\displaystyle V(\phi )\,=\,\frac{\lambda _{n}\,\phi ^{n}}{2}\left( \,1\,+\,%
\frac{\alpha ^{2}\tanh (v_{n}-\phi )}{\beta ^{2}+(\phi -v_{n})^{2}}\right)
\,,  \label{pot1}
\end{equation}
where $\alpha $, $\beta $ and $v_{n}$ are arbitrary constants. We
will take the particular values $n=2$ and $n=4$ from this
potential. The second term controls the bubble nucleation. Its
role is to create an appropriate shape in the inflaton potential,
$V(\phi )$, with a maximum value near $\phi =v_{n}$. The first
term controls inflation after quantum tunneling has occurred, and
its shape coincides with that used in the simplest chaotic
inflationary universe model, $m^{2}\phi {^{2}}/2$. Following
Ref.~\cite{ramon} we take $\alpha ^{2}=0.1$ and $\beta ^{2}=0.01$.
Certainly, this is not the only choice, since other values for
these parameters can also lead to a successful open inflationary
scenario (with any value of $\Omega $, in the range $0 < \Omega <
1$).

We have numerically solved the field equations~(\ref{ecphi})
and~(\ref{ec5}) for the values $ n=2$ and $n=4$ in the effective
potential~(\ref{pot1}). However the instanton has the topology of
a four-sphere, and there are two places at  which $a=0$. These are
the points at which $\tau =0$ and $\tau = \tau_{max}$. Then, the
boundary conditions on $\phi$ arise from the requirement that
$3\dot{\phi}\dot{a}/a$ be finite, i.e $\dot{\phi}(0)=\dot{\phi
}(\tau_{max})=0$. From Eq.(\ref{ec2}), we obtain that at the zeros
of scalar factor, $\dot{a}=\pm 1$.  Since we have  used units
where the Planck mass in four dimension is equal to one, then the
Planck mass in five dimension becomes  $M_{5}\leq
10^{-2}$~\cite{maartens} and by means of Eq.(\ref{mass})  we
arrive to $\sigma = 10^{-10}$.

On the other hand, we take the value of the constant $\lambda
_{n}$ in such a way that an appropriated amplitude for density
perturbation is obtained. Thus, we take the values $\lambda
_{2}=1.5 \times 10^{-6}$ and $\lambda _{4}=10^{-14}$. We choose
$v_{2}=3.5$ and $v_{4}=4.8$, since they provide the needed $60$
e-folds of inflation after tunneling has occurred.

At $\tau \approx 0$, the scalar field $\phi =\phi _{T}$ lies in
the true vacuum, near the maximum of the potential, which (in
Euclidean signature) correspond to $-V(\phi )$. At $\tau \neq 0$,
the same field is found close to the false vacuum, but now with a
different value, $\phi =\phi _{F}$. Specifically, for $n=2$ and
$n=4$ and when the scalar field $\phi $ evolves from some initial
value, i.e. $ \phi _{F}\cong \phi _{i}\approx 3.52$ to the final
value $\phi _{T}\cong \phi _{f}$, numerically we have found that
the CDL instanton does exist, and the brane world open
inflationary universe scenario can be realized. Table
\ref{tab:table1} summarizes our results, which are compared with
those corresponding to Einstein's Theory of Relativity.

\begin{table}
\caption{\label{tab:table1}This table shows the values of $\phi_F$
and $\phi_T$ for which the CDL instanton  exists. }
\begin{ruledtabular}
\begin{tabular}{lll}
Models & $\phi _{F}$
 & $\phi _{T}$  \\ \hline $n=2$ GR
 & $3.52$  & $3.31$  \\ \hline
$n=2$
BW  & $3.52$  & $3.32$  \\
\hline
$n=4$ GR  & $4.85$ & $4.60$ \\
 \hline $n=4$ BW  & $4.88$  & $4.60$  \\
\end{tabular}
\end{ruledtabular}
\end{table}

\begin{figure}[th]
\includegraphics[width=5.0in,angle=0,clip=true]{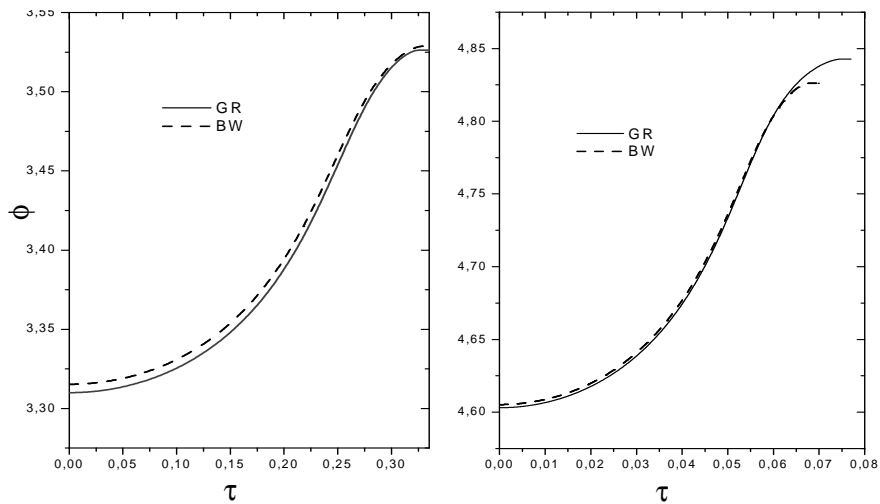}
\caption{The instanton $\phi(\tau)$ as a function of the Euclidean time $%
\tau $ is shown for Einstein GR and  BW. The left panel shows the
case $n=2$, and the  right panel shows the case $n=4$. In both
cases we have taken $\sigma = 10^{-10}$. } \label{fig1}
\end{figure}

Note that the interval of tunneling, specified by $\tau $, decreases when
the $n=4$, but its shapes remain practically similar. The evolution of the
inflaton field as a function of the Euclidean time is shown in Fig.~\ref{fig1}%
.

\begin{figure}[th]
\includegraphics[width=5.0in,angle=0,clip=true]{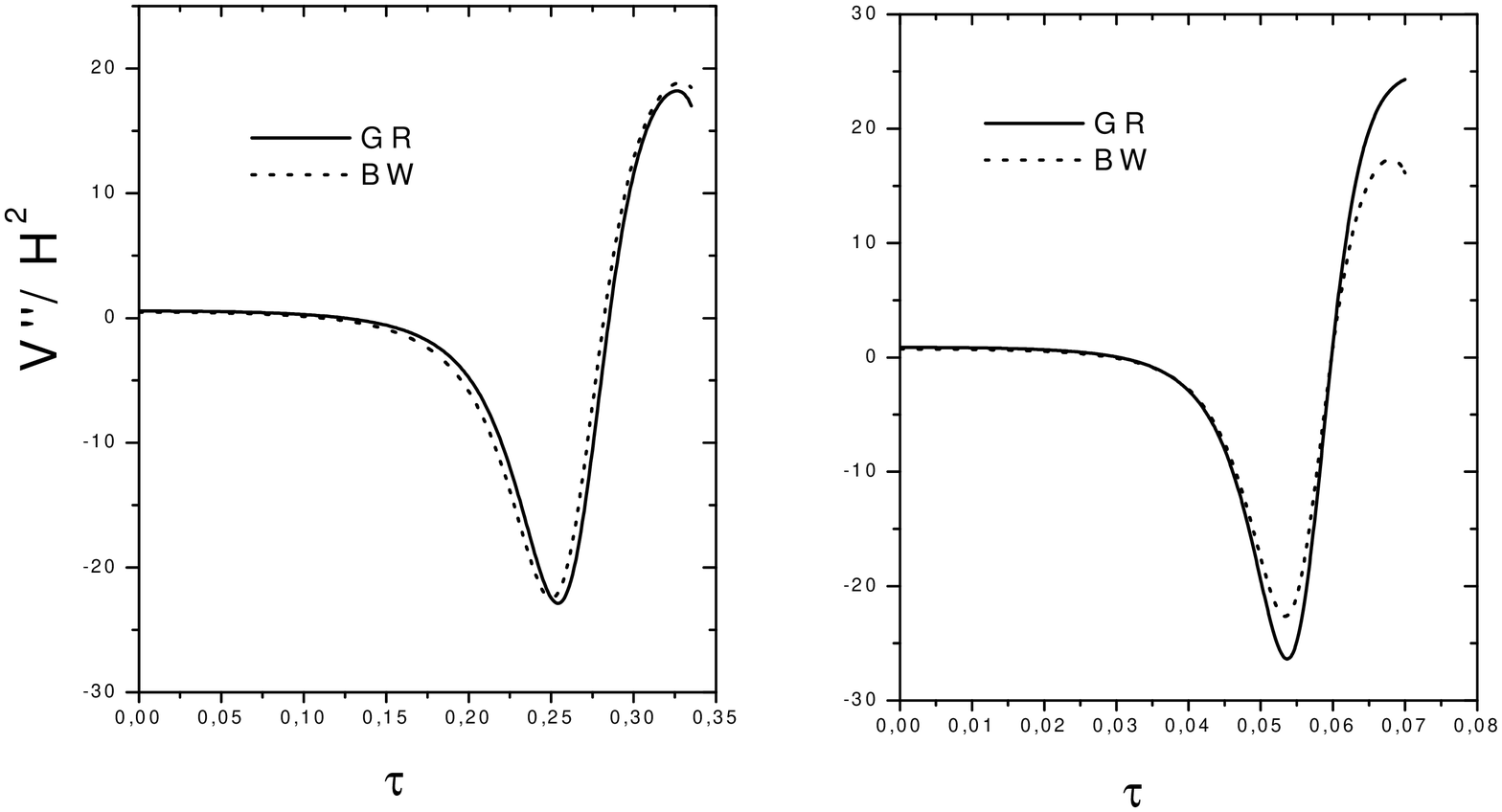}
\caption{This plot shows how during the tunneling process the inequality $%
|V^{^{\prime\prime}}|\gg H^{2}$ holds. The GR lines represents the
same inequality for Einstein's  General Theory of Relativity. The
left panel shows the case $n=2$, and the right panel the cases
$n=4$. In both cases we have taken $\sigma = 10^{-10}$.}
\label{fig2}
\end{figure}

In Fig.~\ref{fig2} we show $|V^{^{\prime \prime }}|\,/\,H^{2}$ as
a function of the Euclidean time $\tau $ for our model. From this
plot we observe that, most of the time during the tunneling, we
obtain $|V^{^{\prime \prime }}|\,>\,H^{2}$, analogous to what
occurs in  Einstein's GR theory. Note that, for  $n=4$, the peak
becomes narrower and deeper, and thus the above inequality is
better satisfied.

It is numerically possible to show that the CDL instanton $%
\phi (\tau )$ exists  for various values of the $\sigma $
parameter. The principal difference shown is the deviation to
general relativity,  but this solution presents a similar behavior
to that described by Linde~\cite{re6}. The values  coincide for
large $\tau $, and its values (after tunneling has occurred)
coincide in the two theories, i.e. Einstein's GR and BW theories.

The following expression represents the instanton action for the
quantum tunneling between the false and the true vacuums in the BW
theory. In order to reproduce the field equations of motions
(\ref{ecphi})-(\ref{ec5}) and using the constraint
(\ref{cosntraint2}), we write for the instanton action

\begin{equation}
\displaystyle S\,=\,2\pi ^{2}\,\int \,d\tau [a^{3}(\frac{1}{2}\phi
^{\prime 2}+V(\phi )+\frac{a^{3}}{2\sigma
}(\frac{1}{2}\phi'^{2}+V(\phi ))^{2}+\frac{3}{\kappa
}(a^{2}\,a''+a\,a'^{2}-a)]. \label{action1}
\end{equation}
We should note that this action is not obtained from
Eq.(\ref{ac1}). At the moment, as far we know nobody has performed
this task. Integrating by parts and using the Euclidean equations
of motion, we find that the action may be written as
\begin{equation}
\displaystyle S\,=\,4\pi ^{2}\,\int \,d\tau \left[ a^{3}\,V(1+\frac{V}{%
2\sigma })+\frac{a^{3}}{8\sigma }\phi'^{4}-\frac{3a}{\kappa
}\right] . \label{bounce}
\end{equation}
Note that this action coincides with that corresponding to its
analogous in the Einstein's general relativity theory if we take
the limit that $\sigma \rightarrow \infty$ \cite{re7}.

The   inflaton field $\phi$ is initially trapped in its false
vacuum, and a  value specified by  $\phi _{F}$ is obtained. After
tunneling to the true vacuum, the instanton gets the value $\phi
_{T}$, and a single bubble is produced . Similar to the case in
the GR theory, the instanton (or bounce) action is given by
$B=S-S_{F}$, i.e. the difference between the action associated
with the bounce solution and the false vacuum. This action
determines the probability of tunneling for the process. We have
defined $V_{F}=V(\phi _{F})$ and $V_{T}=V(\phi _{T})$ as the false
and true vacuum energies, respectively. Under the approximation
that the bubble wall is infinitesimally thin, we obtain the
reduced action for the thin-wall bubble:
\begin{equation}
S=2\,\pi ^{2}\,R^{3}\,S_{1}+\frac{4\pi ^{2}}{\kappa }\left[ \frac{1}{%
H_{T}^{2}}([1-H_{T}^{2}\,R^{2}]^{3/2}-1)-\frac{1}{H_{F}^{2}}%
([1-H_{F}^{2}\,R^{2}]^{3/2}-1)\right] ,  \label{bounce2}
\end{equation}
where we have taken into account the contributions from the wall (first
term) and the interior of the bubble (the second and third terms). Here $R$
is the radius of the bubble, and
\[
H_{F}^{2}=\frac{\kappa \,V_{F}}{3}\left[ 1+\frac{V_{F}}{2\sigma }\right]
\,\,\,\,;\,\,\,H_{T}^{2}=\frac{\kappa \,V_{T}}{3}\left[ 1+\frac{V_{T}}{%
2\sigma }\right] .
\]
The surface tension of the wall becomes defined by
\begin{equation}
\displaystyle S_{1}\,=\,2\,\pi ^{2}\,R^{3}\int \,d\tau \,\left[
\,\phi'^{2}\,\left( 1+\frac{V(\phi )}{\sigma }\right) \right] ,
\label{acc2}
\end{equation}
or
\[
S_{1}\,=\,\,\int_{\phi _{T}}^{\phi _{F}}d\phi [2(V(\phi
)-V_{F})]^{1/2}\left[ 1+\frac{V(\phi )}{\sigma }\right] .
\]

The radius  of curvature  the bubble is one for which the bounce action~(%
\ref{bounce2}) is an extremum. Then, the wall radius is determined
by setting $dS/dR$ = 0, which gives
\[
\frac{S_{1}R\,\kappa }{2}=(1-H_{T}^{2}R^{2})^{1/2}-(1-H_{F}^{2}R^{2})^{1/2}.
\]
This could be solved for the radius of the bubble, and we found
that
\begin{equation}
\displaystyle R=\frac{S_{1}\,\kappa }{\sqrt{\left[ \left( \frac{S_{1}\kappa
}{2}\right) ^{2}+H_{T}^{2}+H_{F}^{2}\right] ^{2}-4H_{F}^{2}H_{T}^{2}}}
\label{r}
\end{equation}
or equivalently
\begin{equation}
\displaystyle R=\frac{S_{1}\,\kappa }{\sqrt{\left[ \left( \frac{S_{1}\kappa
}{2}\right) ^{2}-H_{T}^{2}+H_{F}^{2}\right] ^{2}+H_{T}^{2}(S_{1}\kappa )^{2}}%
}.  \label{rr}
\end{equation}

It is straightforward to check that when $\sigma \rightarrow
\infty $, a correct limit to GR  is obtained. We can introduce a
dimensionless quantity $\Delta \,s$, which represents the strength
of the wall tension in the thin-wall approximationn \cite{MSTTYY}
\begin{equation}
\displaystyle \Delta \,s=\,\frac{S_{1}\,R\kappa }{2}\,<\,1.
\end{equation}
Following the values given in Table 1  and assuming that value
$\sigma =10^{-10}$, we find that the differences in the strength
of the wall tension in the thin-wall approximation becomes $\Delta
\,s_{BW}$, which can be  compared to the corresponding value in
Einstein's GR theory, $\Delta \,s_{GR}$.  For the first model,
i.e, $n=2$, becomes  $\Delta \,s_{BW}-\Delta \,s_{GR}\simeq
\,6.08\cdot \,10^{-3}$. In the second model,  when $n=4$ this
difference becomes on the order of $\Delta \,s_{BW}-\Delta
\,s_{GR}\simeq \,1.01\cdot \,10^{-8}$.  Then, we can see that this
difference in the strength of the wall tension for  the two
theories becomes insignificant.


\section{\label{sec:level3}Inflation after tunneling}

After the tunneling has occurred, we  make an analytical
continuation to the Lorentzian space - time, and we could see what
is the time evolution of the scalar field $\phi (t)$ and the scale
factor $a(t)$. The field equations of motion for the fields $\phi
$ and $a$ are given by
\begin{equation}
\displaystyle \ddot{\phi}+3\frac{\dot{a}}{a}\dot{\phi}+V_{,\phi }(\phi
)=0\,\,,  \label{ec8}
\end{equation}
and
\begin{equation}
\displaystyle \frac{\ddot{a}}{a}=-\frac{8\pi }{3M_{4}^{2}}\left(\dot{\phi}%
^{2}-V(\phi )+\frac{1}{8\sigma }(5\dot{\phi}^{2}-2V(\phi ))(\dot{\phi}%
^{2}+2V(\phi ))\right) ,  \label{ec9}
\end{equation}
where the dots now denote derivatives with respect to the cosmological time.

In order to solve this set of Equations numerically, we use the
following boundary conditions: $\dot{\phi}(0)=0$, $a(0)=0$ and $\dot{a}%
(0)=1$. As in  \cite{re6}\cite{shiba} and \cite{ramon}, the scalar
field slowly rolls down to its minimum of effective potential, and
its field starts to oscillate near this minimum. During this stage
the $N$ e-folds parameter presents different values for our models
under study; those results are summarized in Figure 3. Clearly,
the $N$ e-folds parameter increases in the BW scenarios.
\begin{figure}[th]
\includegraphics[width=5.0in,angle=0,clip=true]{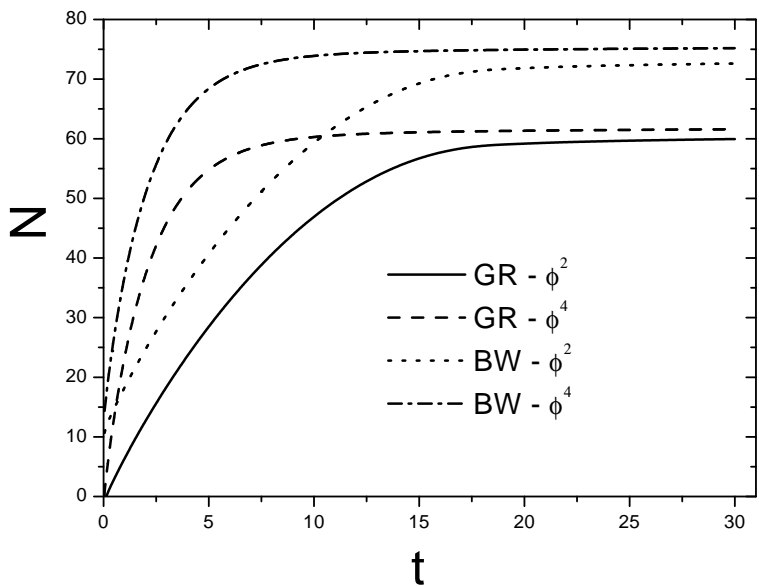}
\caption{This plot shows the number $N$ the e-folds as a function
of the cosmological times $t$. The GR line represents Einstein's
GR theory. In this graph we have assumed that the constant
$\sigma=10^{-10}$. } \label{fig3}
\end{figure}

\section{\label{sec:level4}Spectrum of scalar perturbation }

Even though the study of scalar density perturbations in open
universes is quite complicated~\cite{re8}, it is interesting to
give an estimation of the standard quantum scalar field
fluctuations inside the open bubble. The corresponding density
perturbation in the brane world cosmological model becomes
\cite{maartens}

\begin{equation}
\frac{\delta \rho }{\rho }\approx Cte*\left( \frac{V}{V^{\prime }}\right) ^{%
\frac{3}{2}}\left( \frac{2\sigma +V}{2\sigma }\right) ^{\frac{3}{2}},
\label{pertur}
\end{equation}
where $Cte=\frac{24}{5}\sqrt{\frac{8\pi }{3}}$. Note that the
latter equation coincides with its analogous Einstein Eqs., when
the limit  $\sigma \rightarrow \infty$ is taken. Certainly, other
contributions must be added in order to get an exact expression
\cite{re6,re8}, but those contributions do not change expression
(\ref{pertur}) significantly if we use it for $N>3$.

\begin{figure}[th]
\includegraphics[width=3.0in,angle=0,clip=true]{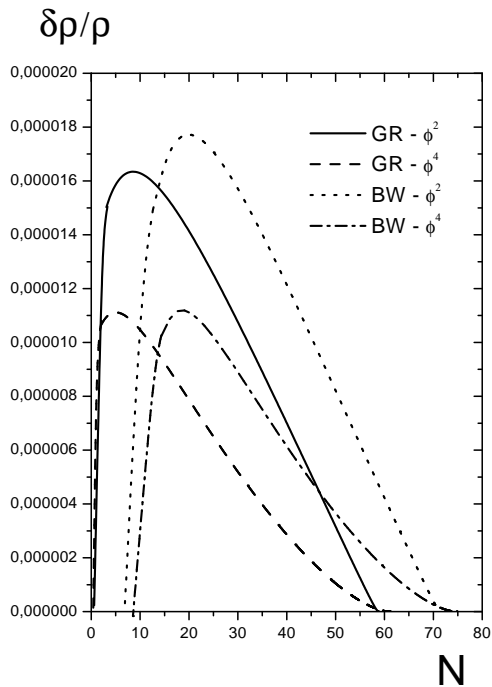}
\caption{Scalar density perturbations for our models as a funtion
of N e-folds. These plots are compared with those obtained by
using the Einstein's GR theory, where
$\delta\rho$/$\rho\approx\,CteH^{2}$/$|\dot{\phi}|$.} \label{fig4}
\end{figure}

Figure~\ref{fig4} shows the magnitude of the scalar perturbations
$\delta \rho /\rho $ for our models as a function of the $N$
e-folds of inflation, after the open universe was formed. Even
though the shape of the graph is similar to that of Einstein's GR
theory , the maximun value of $\delta \rho /\rho $ has become
bigger in the BW then in Einstein's GR models. For instance, in
Einstein's  GR they it becomes maximum for $N\sim\,0(12)$, while
for BW model its maximum  is found at $N\sim\,0(20)$. Also in the
the model with $n=4$, the values of $N$ where $\delta \rho /\rho $
vanishes, becomes bigger in BW than in Einstein's GR theory. On
the other hand, we should mention that there is relationship
between the values of $N$ and the scale where $\delta\rho/\rho$ is
measured.  For $N\sim 10$, where $\delta \rho /\rho $ gets it
maximum value, it is found that the scale where the scalar
perturbation is measured corresponds to the $10^{24}cm$. However,
for $N\sim 15$ it decreases to $10^{22}cm$, and for $N\gg 50$ this
practically comes to zero. Something similar happened with for the
case $n=4$. There, the corresponding values of $N$  were smaller.

Also, it is interesting to give an estimation of the tensor
spectral index $n_T$  in the brane world cosmological model. Using
Ref.\cite{maartens}, this index  for a flat universe is given by
\be n_T\simeq\,-\frac{1}{8\pi}\left(\frac{V\,'}{V}\right)^2\,
\left[1+\frac{V}{2\sigma}\right]^{-1}, \en

By numerically solving the field equation associated with field
$\phi$, we obtain for the cases $n=2$ and $n=4$ in the GR theory
the following values: $n_T\simeq\,-0.0350$ which is evaluated in
the value the $N$, where $\delta\rho/\rho$ presents a maximum,
i.e, $N\simeq9$ and $n_T\simeq\,-0.0610$ for $N\simeq5$. In BW
cosmological models for the cases $n=2$ and $n=4$, we obtain
$n_T\simeq-0.0315$ for $N\simeq20$ and $n_T\simeq-0.0611$ for
$N\simeq19$.

\section{conclusion}

In this paper we have studied one-field open inflationary universe
models in which the gravitational effects are described by BW
cosmology. In this kind of theory, the Friedmann equation gets
modified by an additional  term: $\rho^2/2\sigma$. We have
solutions to an effective potentials in which the CDL instantons
exist \cite{ramon}. The existence of these instantons becomes
guaranteed since the inequality $|V^{''}|>H^{2}$ is satisfied, and
thus, with the slow-roll approximation, inflationary universes
models are realized for different values of the parameter $n$. For
the two values  ($n=2$ and $n=4$) that we considered, $V^{''}$
remains greater than $H^{2}$ during the first e-folds of
inflation.

 Also, we have generalized the CDL instanton
action to brane world cosmology. This action is described by
expression (\ref{action1}).

On the other hand, it seems that, according to equations
(\ref{bounce2}) and (\ref{acc2}),  the result for the probability
of nucleation of a bubble is the same as in Einstein gravity.
However, this is superficial, since there is a modified
relationship between the Hubble rate and the potential $V(\phi)$,
given by the well known modified Friedmann equation and thus a
modified expression between the wall tension $S_{1}$ and the
potential $V(\phi)$ occurs. In the thin-wall limit, we have found
that the strength of the wall tension, -$\triangle\, s_{BW}$-
minimum increase, when compared with their analogous results
obtained in Einstein's GR theory.

We have also found that the inclusion of the additional term
($\rho^2$) in the Friedmann's equation improves some of the
characteristic parameters of inflation. For instance, this is
accentuated in the number $N$ the e-folds (see Fig.[\ref{fig3}]).

Finally, in the $\delta\rho/\rho$ graphs the maximum presents a
displacement when compared with that obtained in Einstein's GR
theory, this would change the value of the fundamental parameter
$\lambda_n$ that appears in the scalar potentials.

\begin{acknowledgments}
S.d.C.and J.S was supported from COMISION NACIONAL DE CIENCIAS Y
TECNOLOGIA through FONDECYT Grant Nos. 1030469; 1010485 and
1040624 and Postdoctoral Grant 3030025. Also, it was partially
supported by UCV  Grant No. 123.752. R. H. is supported from PUCV
through Proyecto de Investigadores J\'ovenes a$\tilde{n}$o 2004.
The authors wish to thank CECS for its kind hospitality.
\end{acknowledgments}

\end{document}